\documentclass[reprint,
]{revtex4-1}

\usepackage{amsmath}
\usepackage{amssymb}
\usepackage{listings}
\usepackage{graphicx}
\usepackage{dcolumn}
\usepackage{bm}
\usepackage{natbib}
\usepackage[colorlinks,citecolor=red,linkcolor=blue,urlcolor=blue]{hyperref}
\usepackage{caption}
\usepackage{hhline}
\usepackage{float}
\usepackage{mhchem}
\usepackage{calrsfs}
	\DeclareMathAlphabet{\pazocal}{OMS}{zplm}{m}{n}
\usepackage{physics}
\usepackage[toc,page]{appendix}
\usepackage[export]{adjustbox}

\newcommand{\mn}{m\boldsymbol{\alpha}_1+n\boldsymbol{\alpha}_2}
\newcommand{\pq}{p\boldsymbol{\alpha}_1+q\boldsymbol{\alpha}_2}
\newcommand{\squeeze}[2]{\left\lbrace #1 \middle| #2\right\rbrace}
\newcommand{\squeezeA}[2]{\left\lbrace #1 , #2\right\rbrace}
\newcommand{\bk}{\bold{k}}

\newcommand{\capk}[1]{\hat{#1}_{\bold{k}}}
\newcommand{\capmk}[1]{\hat{#1}_{-\bold{k}}}

\newcommand{\capdagk}[1]{\hat{#1}^\dagger_{\bold{k}}}

\newcommand{\expk}[1]{e^{i\bold{k}\cdot\bold{#1}}}
\newcommand{\expmk}[1]{e^{-i\bold{k}\cdot\bold{#1}}}

\newcommand{\done}[0]{\delta_1}
\newcommand{\dtwo}[0]{\delta_2}

\newcommand{\ddY}[1]{\frac{\partial }{\partial #1}}

\newcommand{\uu}[1]{u^{#1}(\bold{k})}

\newcommand{\squeezeB}[3]{\left\langle #1\middle| #2\middle| #3\right\rangle}

\newcommand{\squeezeD}[3]{\left\langle #1\middle| #2\middle| #3\right\rangle}

\newcommand{\dXdY}[2]{\frac{\partial #1}{\partial #2}}

\begin{document}


\title{The topological magnon bands in the Flux state in Sashtry-Sutherland lattice}

\author{Dhiman Bhowmick}
\author{Pinaki Sengupta}%
\affiliation{%
 School of Physical and Mathematical Sciences, Nanyang Technological University, 21 Nanyang Link, Singapore 637371, Singapore \\
}%

\date{\today}

\begin{abstract}
We investigate low energy magnon excitations above the non-collinear flux state and
non-coplanar canted flux state in a Heisenberg anti-ferromagnet with 
Dzyaloshinskii-Moriya interaction~(DMI) on a Sashtry-Sutherland lattice.
While previous studies have shown the presence of topological magnetic excitation 
in the dimer and ferromagnetic phases on the Shastry-Sutherland lattice, 
our results establish the non-trivial topology of magnons in the anti-ferromagnetic 
flux and canted flux states. Our results uncover the existence of a multitude
of topological phase transitions in the magnon sector -- evidenced by the changing
Chern numbers of the single magnon bands -- as the Hamiltonian parameters are varied, even when the ground state remains unchanged. The thermal Hall conductivity
is calculated and its derivative is shown to exhibit a logarithmic divergence at the
phase transitions, independent of the type of band touching involved. This may provide
a useful means to identify the energy at which the transition occurs. Finally, we propose the way to realize the studied model in a practical material.
\begin{description}
\item[PACS numbers]
 
\end{description}
\end{abstract}

\pacs{Valid PACS appear here}
\maketitle


\section{\label{sec1}Introduction}
The study of topological phases of matter has gained widespread interest
during the past decade. While topological phases are realized in both fermionic\cite{Fermionic1,Fermionic2}
and bosonic systems\cite{bosonic1,bosonic2,bosonic3,bosonic4}, much of the advancement (theoretical investigations and experimental realizations of topological phases) has been confined to
fermionic systems; study of bosonic topological
phases have attracted widespread interest relatively recently. This is partly due to the 
fact that topological character of ground state phases of (non-interacting) fermions is readily
identified from the properties of the energy bands and there exist 
well developed experimental probes to detect them. On the other hand, 
the ground state of bosonic systems is often a condensate and topological
character is manifested in low lying excitations\cite{bosonic1,bosonic3,bosonic4}. 

Quantum magnets have served as a versatile test bed for realizing novel bosonic
phases, including bosonic topological phases. The topological character of the
magnetic phase is manifested through the behavior of magnons.
Magnons are charge neutral quasi-particle excitations in insulating magnetic 
systems. Analogous to electrons in standard topological insulators, magnons in 
magnetic insulators exhibit thermal Hall effect\cite{THEIntro}, spin-Nernst 
effect\cite{SNEIntro1,SNEIntro2,SNEIntro3,SNEIntro4}, and magnon-driven spin Seebeck 
effect\cite{SpinSeebackIntro}. The interest in these systems is driven by both
fundamental reasons and potential for technological applications. The recent use
of Skyrmions in spintronics for efficient magnetic storage and read/write 
devices with minimal Joule heating effect\cite{Skyrmion1,Skyrmion2} underscores
the potential practical applications of topologically no-trivial magnetic states. 
The wide range of quantum magnets with 
varying interactions and lattice structures as well as the ability to control
the number of quantized excitations with an external magnetic field make them 
ideal for exploring novel magnetic phases. Geometrically frustrated quantum magnets 
are particulalry promising in realizing and controlling topologically non-trivial
spin textures\cite{frustration1,triangular, FluxStateIntro,frustration2,frustration3,frustration4}. The interplay between competing interactions, geometric frustration
and external magnetic field result in a wide variety of magnetic phases that are
not commonly observed in their non-frustrated counterparts. In most cases, topological 
excitations in quantum magnets are driven  Dzyaloshinskii-Moriya 
interaction~(DMI)\cite{CoplanarIntro1,CoplanarIntro2, 
CoplanarIntro3,CoplanarIntro4,CoplanarIntro5}, although topological magnon bands can 
exist without DMI as well~\cite{NonCoplanar1,NonCoplanar2} due to non-coplaner chiral spin-texture. Most strikingly, the change 
in spin-texture by changing the parameters in the non-coplanar spin systems, gives rise 
to variety of topological phases in the same system\cite{NonCoplanar2,NonCoplanar3}.

The Shastry-Sutherland model is a paradigmatic model for the study of frustrated
magnetism. Since the degree of frustration can be tuned by varying the ratio of the 
diagonal and axial bonds, the model exhibits a wide range of novel magnetic phases\cite{phase1,phase2,phase3}.
The existence of a number of materials with underlying SS geometry of the magnetic ions offers the prospect
of observing theoretically predicted phases and phenomena in real materials\cite{MaterialRealization,RB41,RB42,RB43}. Since 
DMI is ubiquitous in all of these materials, it is natural to supplement the 
canonical SS model with DMI. This results in an even richer variety of magnetic
orderings including colinear, coplanar and non-coplanar spin configurations, several of 
which host topological magnons\cite{FluxStateIntro,Shahzad_2017}. Previous studies of topological magnons for the 
SS lattice were restricted to the dimer~\cite{triplon} and the ferro-magnetic 
phases~\cite{ferroIntro}. In recent past there has been a growing interest in
studying topological magnons in non-collinear spin configurations in frustrated
lattices. Here we present the results of our investigation of topological magnons
in the recently proposed flux state, that is stabilised by DMI perpendicular to the lattice plane  in the SS 
lattice\cite{FluxStateIntro}.In presence of in plane DMI, 
this evolves to the canted flux state -- the resulting magnon bands in an external longitudinal magnetic field, carry non-zero
Chern numbers that determine the topological character of the magnon bands. Varying
the different components of the DMI result in a sequence of topological phase
transitions where the Chern number for the magnon bands change over a wide range of
possible values. 


The topological phase transitions in the system can be detected from the first derivative of the thermal Hall conductivity which exhibits a logarithmic divergence at the  transition\cite{NonCoplanar2,divergence}. The peak height of the logarithmic divergence increases with temperature following an  algebraic relation. The interpolation of the first derivative of the thermal Hall conductance as a function of temperature yields information on the nature of band touching (gap closing) at the phase transition.

\section{\label{sec2}Model Hamiltonian and Method}
\begin{figure}[H]
	\centering
		\includegraphics[width=0.3\textwidth]{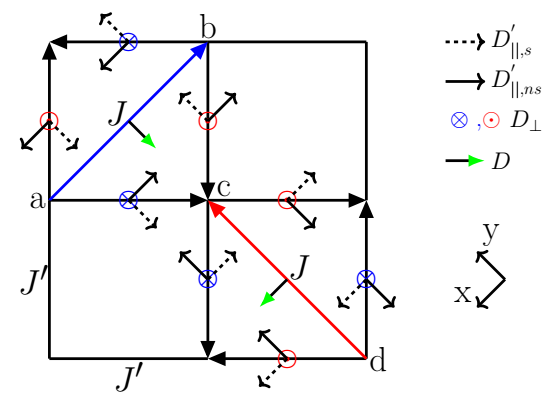}
	\caption{(color online) The Heisenberg and Dzyaloshinskii–Moriya(DM) interactions on Shastry-Sutherland lattice.}
	\label{lattice}
\end{figure}

The model Hamiltonian is given by,
\begin{align}
\pazocal{H}=&J\sum_{\left\langle\left\langle i,j \right\rangle\right\rangle} \bold{S}_i\cdot\bold{S}_j + J' \sum_{\left\langle i,j \right\rangle} \bold{S}_i\cdot\bold{S}_j \nonumber\\
  &+ \bold{D}\cdot\sum_{\left\langle\left\langle i,j \right\rangle\right\rangle} \left(\bold{S}_i\times\bold{S}_j\right)+ \bold{D'}\cdot\sum_{\left\langle i,j \right\rangle} \left(\bold{S}_i\times\bold{S}_j\right) \nonumber \\
  &-B\sum_i\bold{S}_i^z,
  \label{eq1}
\end{align}
where $J$ and $J'$ ($>0$)  are anti-ferromagnetic Heisenberg spin-exchange  on the 
axial and diagonal bonds of the SS lattice respectively (henceforth referred to as SS 
bonds).   $\bold{D}$ and $\bold{D'}$ are DM-vectors on the SS bonds as shown in Fig.
\ref{lattice}. The nature of DM interactions are chosen according to the symmetry
constraints of the SS lattice~\citep{triplon,triplon2,triplon3}. The origin of the 
DM interactions are further discussed in the section Sec.\ref{sec5}.

We start with a classical ground state and investigate 
quantized low energy excitations, magnons, focusing on identifying any 
topological character. The classical ground state of the 
Hamiltonian Eq.\ref{eq1} is derived by replacing the local spin moments by classical vectors of 
unit magnitude. The state of each spin is specified by the polar co-ordinates,
\begin{equation}
\bold{S}=S(\sin(\theta)\cos(\phi), \sin(\theta)\sin(\phi),\cos(\theta)).
\label{eq2}
\end{equation} 
The ground state spin configuration is obtained by minimizing the energy of the 
Hamiltonian w.r.t the angles $\theta$ and $\phi$. For the ground state phases of interest, 
viz., the flux and the canted flux states, the magnetic unit cell is of the same size as the unit cell of the SS lattice and consists of four sites 
as shown in Fig.\ref{lattice}. 

The classical phases are further discussed in Sec.\ref{sec4}. Since we are interested
in primarily the flux and the (in plane DMI induced) canted flux states, we start by 
identifying the parameter ranges where these are realized. The ground state phases of the
Hamiltonian Eq.(\ref{eq1}) for classical spins has been investigated in Ref.[\onlinecite{FluxStateIntro}]
-- with no in-plane component of DMI, the flux state (Fig.\ref{FluxState}(a)) is 
stabilized above a critical value of the normal component of the DMI along the axial bonds, $D_\perp$(Fig.\ref{lattice}) (the DMI on the diagonal bonds are constrained by the symmetry
of the lattice to lie on the plane of the lattice). The continuous U(1) symmetry of the Hamiltonian is spontaneously broken in the flux state and Fig.\ref{FluxState}(a) shows one of the degenerate ground states. Interestingly, the flux state state is also realized in 
the square lattice, but for a much stronger DMI. The geometric frustration of the SS lattice
facilitates the appearance of the flux state for a more moderate (and realistic) strength
of DM interaction. However, the symmetry of the SS lattice allows for in-plane DMI components of DMI, denoted in this work by $D$ on the diagonal bonds and $D_{||,s}$, $D_{||,ns}$ on the axial bonds (see Fig.\ref{lattice}). Any non-zero in-plane DMI tilts the spins out of plane keeping the in plane spin component of nearest-neighbour sites perpendicular to each other. The spins on the two distinct diagonal bonds cant in opposite
direction -- for one of the diagonals, the spins cant out of the plane, whereas for the other diagonal, they cant into the plane of the lattice. The in-plane components are aligned along the diagonal bonds, as depicted in Fig.\ref{3Q-order}(a). This ground state spin configuration is referred to
as the canted-flux state. In the presence of the in-plane components of DMI, the U(1)
symmetry of the Hamiltonian is explicitly broken and there is no spontaneous breaking of U(1) symmetry in the in-plane DMI driven canted flux state.

To study the excitations above magnetic ground state, we have used the linearized 
Holstein-Primakoff transformation\cite{bosonic1,Owerre_2017}. The Holstein-Primakoff transformation is a  versatile
and extensively used approach to study low energy magnon excitations above magnetically
ordered ground state phases in quantum magnets.
Previous implementations of this method have largely been restricted to collinear magnetic orderings~\cite{bosonic1, ferroIntro, bosonic3,CoplanarIntro4,CoplanarIntro3,paramagnet,StripeZigzag}
although there have been recent attempts to extend it to 120$^\circ$ non-collinear magnetic
order in triangular lattices~\cite{triangular}. In this work, we have extended the 
Holstein-Primakoff approach to study magnon excitations above complex magnetic
orders with longer periodicity. Here we present a brief discussion the method. First, the local co-ordinate axis at each site of the lattice is rotated such that the $S_z$ axis is aligned along the local spin direction. For low temperature excitations the linearized Holstein-Primakoff transformation is given by,
$
\hat{S}'^+_{i,a}=\sqrt{2S} \hat{a}_i,\;
\hat{S}'^-_{i,a}=\sqrt{2S} \hat{a}^\dagger_i ,\;
\hat{S}^z_{i,a}=S-\hat{a}^\dagger_i\hat{a}_i,
$
where, we consider $\hbar=1$ and $\hat{a}_i^\dagger (\hat{a}_i)$ represent creation (annihilation)
operators for quantized excitations above the magnetic ground state at site $i$. These obey bosonic
commutation relations,
$
[\hat{a}_i,\hat{a}_j^\dagger]I=\delta_{i,j}
$
and 
$ 
[\hat{a}_i,\hat{a}_j] = 0 = [\hat{a}_i^\dagger,\hat{a}_j^\dagger] 
$. Since the unit cell consists of 4 sites, there are four species of bosons corresponding
to each inequivalent lattice site. In the next section, the detailed classical ground state and corresponding magnon bands and their topological properties are discussed.


\section{\label{sec4} Results and Discussion}
\subsection{Flux State}

\begin{figure}[H]
	\centering
		\includegraphics[width=0.5\textwidth]{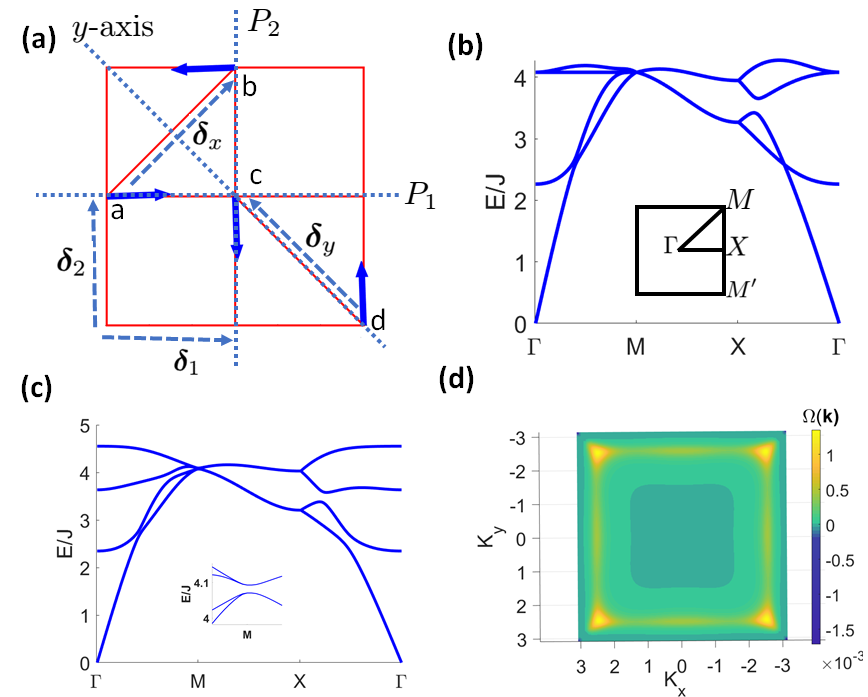}
	\caption{(color online) (a) The flux state in Shastry-Sutherland lattice. (b) The magnon band structure for $J=1.0, J'=1.1, D_\perp=0.8, B_z=0.0$. The inset of the figure shows the Brillouin zone and high symmetry lines. (c) The magnon band structure for $J=1.0, J'=1.1, D_\perp=0.8, B_z=0.5$. Inset of the figure shows the magnified magnon band structure to show the lifting of four-fold degeneracy at M-point. (d) The non-abelian Berry curvature of lowest and second-lowest bands for $J=1.0, J'=1.1, D_\perp=0.8, B_z=0.5$. For convenience the negative Berry-curvature which is concentrated at M-point is divided by 100, to increase the visibility of the Berry-curvature distribution throughout the Brillouin zone.}
	\label{FluxState}
\end{figure}

The flux state is stabilized as the ground state when the component of the DMI 
normal to the plane of the lattice on the axial bonds, $D_\perp$, exceeds a critical magnitude (e.g. when $J\approx J'$, flux state is stable for $D_\perp\gtrapprox 0.6J$) and the in-plane components vanish for all DMI (the DMI on the diagonal bonds are constrained by symmetry requirements to be strictly in-plane).
The flux state is comprised of  the nearest-neighbour spins aligned perpendicular to each other and parallel
to the plane of the lattice (Fig.\ref{FluxState}(a)), which is energetically favored by the 
perpendicular DM-component $D_\perp$. The state is characterized by the spontaneous breaking of continuous U(1) spin-rotation symmetry about the z-axis. In the bosonic (magnon) language, the ground state (flux state) is the vacuum and the Holstein-Primakoff bosons represent quantized low energy excitations above this ground state. The magnon bands at zero magnetic field are shown in 
the Fig.\ref{FluxState}(b). At the $\Gamma$ point the lowest band becomes gapless 
revealing the presence of Goldstone-mode associated with U(1) symmetry breaking. The bands along line-$\overline{\text{MX}}$ are twofold degenerate -- these can be understood in terms of Kramer's degeneracy\cite{StripeZigzag,Symmetry}. The operator $\hat{m}_2=\left\lbrace \tilde{M}_2 \tau|\boldsymbol{\delta}_2\right\rbrace$ commutes with the Hamiltonian where $\tilde{M}_2$ is the reflection operator along the $P_2$ axis
and $\boldsymbol{\delta}_2$ is the translation by have lattice parameter along the
same axis, as shown in Fig.\ref{FluxState}(a); $\tau$ is the time-reversal operator. But, $\hat{m}_2$ is not the symmetry operator for the classical ground state shown in the Fig.\ref{FluxState}(a). Instead,  symmetry operator for the ground state is given by $\hat{m}'_2=\left\lbrace \tilde{M}_2 e^{i\pi\hat{S}^y} \tau|\boldsymbol{\delta}_2\right\rbrace$, which contains an additional rotation of spin by $\pi$ about the $y-$axis. On the line-$\overline{\text{MX}}$ in the Brillouin zone, $(\hat{m}'_2)^2=e^{ik_x}=-1$. Hence $\hat{m}'_2$ is anti-unitary operator with a squared value of $-1$, which in turn, results in the Kramer's degeneracy. $\hat{m}'_2$ maps one Kramer's degenerate wavefunction along the  $\overline{\text{MX}}$-line to the other Kramer's degenerate wavefunction along the $\overline{\text{M'X}}$-line. Further, the symmetry operation $\hat{c}_{2,1}=\left\lbrace \tilde{C}_{2,1} e^{i\pi\hat{S}^y} |\boldsymbol{\delta}_1\right\rbrace$, maps the Kramer's degenerate state from $\overline{\text{M'X}}$ to $\overline{\text{MX}}$, where $\tilde{C}_{2,1}$ is the two-fold rotation about the $P_1$ axis and $\boldsymbol{\delta}_1$ as shown in the Fig.\ref{FluxState}(a). Thus, the band degeneracy along $\overline{\text{MX}}$-line is protected by symmetries $\hat{m}_2$ and $\hat{c}_{2,1}$. Additionally, there is four-fold degeneracy at the M-point due to the presence of the symmetry $\hat{m}_y=\left\lbrace \tilde{M}_y\tau|\bold{0}\right\rbrace$, where $\tilde{M}_y$ is the reflection operator about the $y-$axis as shown in Fig.\ref{FluxState}(a). The symmetry $\hat{m}_y$ 
maps one pair of Kramer's degenerate state to the other pair of Kramer's degenerate state at M-point. The Berry-curvature is not well defined for bands, because the bands are degenerate.

The band structure in the presence of an external longitudinal magnetic field is shown in Fig.\ref{FluxState}(c). Application of magnetic field produces a finite out of plane spin component, which breaks the $\hat{m}_y$-symmetry. Thus, the four-fold degeneracy reduces to two fold degeneracy at the M-point,  as shown in the inset of Fig.\ref{FluxState}(c). But the band sticking along $\overline{\text{MX}}$-line and Goldstone modes at $\Gamma$-point are preserved. 
The degeneracy of the bands prevents the calculation of the standard Berry curvature 
and Chern numbers of the bands, although there is no symmetry constraints to make 
Berry curvature zero. One can, however, calculate the non-abelian Berry-curvature and non-abelian Chern number of the degenerate bands following the procedure outlined in Ref.\onlinecite{Hatsugai}(see Appendix.\ref{AppendixE} for details). The results for the non-abelian Berry curvature of the lowest and second-lowest band is shown in Fig.\ref{FluxState}(d). The negative part of Berry-curvature is highly concentrated at the M-point and the positive berry curvature is distributed in the remaining Brillouin zone. The inversion symmetry $\hat{i}$ of the system with an inversion center at point-c in Fig.\ref{FluxState}(a) is also reflected in the Berry-curvature. 
Our results show that the negative and positive contribution of the non-abelian Berry curvature cancels, yielding a vanishing non-abelian Chern number for the lower (or the upper) pair of bands. 

\subsection{Canted flux state}
\subsubsection{\textbf{Topological magnon bands and topological phase diagram}}
\begin{figure}[H]
	\centering
		\includegraphics[width=0.5\textwidth]{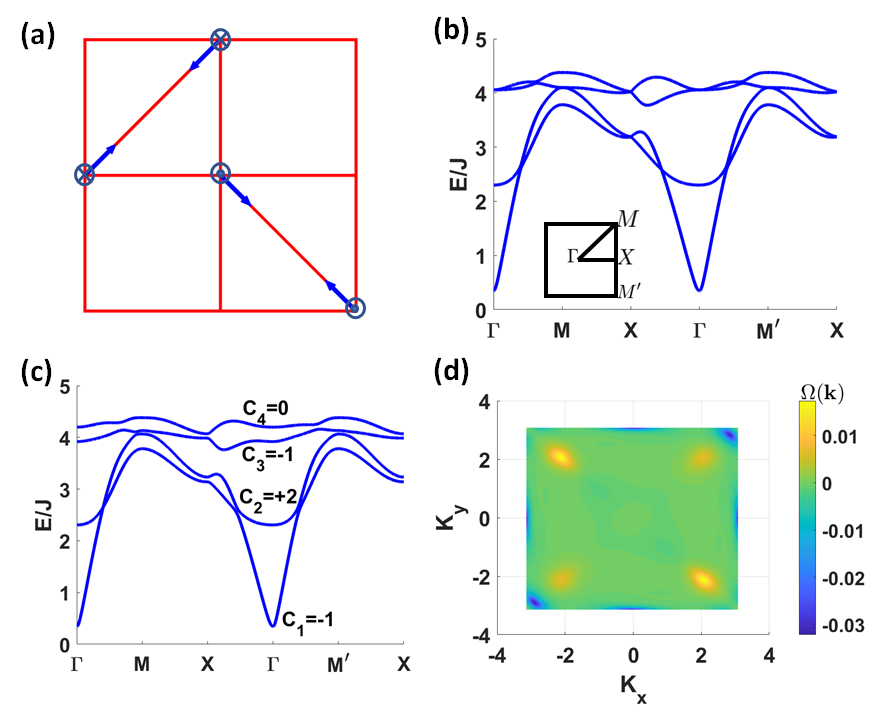}
	\caption{(color online) (a) The spin configuration in 3Q-ordered phase. Circle with dot and the circle with cross represent spin component out of plane upward and downward directions respectively. (b) The magnon band structure for $J=1.0, J'=1.1, D_\perp=0.8, D=0.2, D_{||,s}=0.05, D_{||,ns}=0.1, B_z=0.0$. The inset of the figure shows the Brillouin zone and high symmetry lines and high symmetry points. (c) The magnon band structure for $J=1.0, J'=1.1, D_\perp=0.8, D=0.2, D_{||,s}=0.05, D_{||,ns}=0.1, B_z=0.3$. The Chern numbers of the bands are also shown in the Figure. (d) The Berry curvature third Band for $J=1.0, J'=1.1, D_\perp=0.8, D=0.2, D_{||,s}=0.05, D_{||,ns}=0.1, B_z=0.3$.}
	\label{3Q-order}
\end{figure}

When in-plane components of the DM-interactions (both along the axial as well as the diagonal bonds), a canted-flux state, as shown in the Fig.\ref{3Q-order}(a), is
realized as the ground state. The in-plane components of spins are directed along the diagonal bonds; additionally, the spins acquire an out-of-plane component. There is two-fold degeneracy in the ground state configuration -- one of them is shown in Fig.\ref{3Q-order}(a) with the spins pointing inwards along the diagonal bonds; the other degenerate state is obtained by flipping the spins so that they point outward along the diagonal bonds and the out-of-plane spin components are also flipped. The spins are canted away of the plane of lattice at an angle given by ,
\begin{equation}
\theta=\frac{\pi}{2}+\frac{1}{2}\tan^{-1}\left(\frac{4D+8(D_{||,s}-D_{||,ns})}{4J+8D_\perp-8J'}\right),
\end{equation}
where the spins along one diagonal are canted out of the plane while those along the 
other diagonal are canted into the plane. 

 The magnon bands for the canted-flux state is shown in the Fig.\ref{3Q-order}(b), where the bands are observed to be degenerate at $\Gamma$, M and X-point in the Brillouin zone. These degeneracies are protected by symmetries(See Appendix.\ref{AppendixC} for details).

In the presence of an external longitudinal magnetic field, the canting angles of the 
two pairs of spins are no longer identical. The energy of the canted flux state in a
magnetic field is given by,
\begin{align}
\frac{E_{cl}}{NS^2}=&J(cos(2\theta_1)+\cos(2\theta_2))+J'\cos(\theta_1)\cos(\theta_2)
\nonumber\\
&+D(\sin(2\theta_1)-\sin(2\theta_2))-8D_\perp\sin(\theta_1)\sin(\theta_2)
\nonumber\\
&+4(D_{||,s}-D_{||,ns})\sin(\theta_2-\theta_1)\nonumber \\
&-\frac{2B}{S}(\cos(\theta_1)+\cos(\theta_2)),
\label{eq6}
\end{align}
where $\theta_1$ and $\theta_2$ are the (different) canting angles angles made by the spins into
and out of the plane of the lattice. The energy is minimized for $\theta_1\neq \theta_2$
-- the application of magnetic field not only lifts the non-symmorphic symmetries 
but also renders the magnetic symmetry group of the ground state trivial. As a consequence, all the four bands are gapped out, as shown in Fig.\ref{3Q-order}(c). The magnon Hamiltonian in presence of all DM-interactions and magnetic field can be found in the Appendix.\ref{AppendixD}.

The lifting of the symmetry-protected degeneracies in the energy bands allows us to
calculate the Berry curvature and Chern number associated with each band separately
in the usual manner(see Appendix.\ref{AppendixE}). 
The results reveal that the energy bands acquire topological character for the
parameter set chosen in Fig.\ref{3Q-order}. A representative Berry-curvature 
distribution for the third band is shown in Fig.\ref{3Q-order}(d). The Berry
curvature is concentrated near the M and M' points of the Brillouin zone. The 
existence of the two-fold rotational symmetry at the center of the diagonal bond 
is reflected in the Berry-curvature. Unlike the flux state, the Berry curvature across 
the entire Brillouin zone do not cancel and this results in a non-zero Chern number
for three of the four bands (Fig.\ref{3Q-order}(c))(see Appendix.\ref{AppendixE} for details).

\begin{figure}[H]
	\centering
		\includegraphics[width=0.5\textwidth]{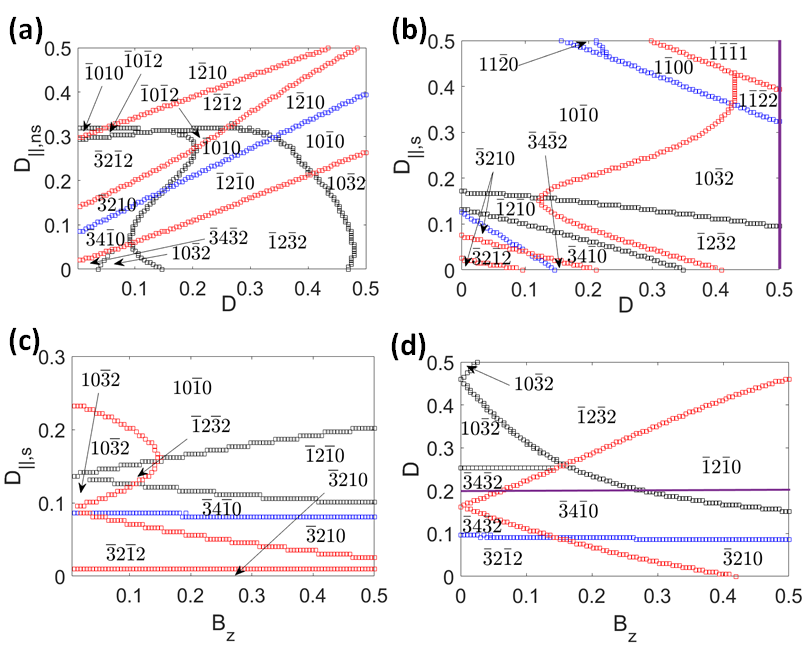}
	\caption{(color online) (a) The topological phases of magnon in the parameter space $D$ and $D_{||,ns}$, with $D_{||,s}=0.1, B_z=0.2$. (b)The topological phases of magnon in the parameter space $D_{||,s}$ and $D$, with $D_{||,ns}=0.1, B_z=0.2$. (c) The topological phases of magnon in the parameter space $D_{||,s}$ and $B_z$, with $D_{||,ns}=0.1, D=0.05$. (d) The topological phases of magnon in the parameter space $D$ and $B_z$, with $D_{||,s}=0.05, D_{||,ns}=0.1$. For all the plots the residual parameters are taken as $J=1, J'=1.1, D_{\perp}=0.8$. The four digits from left to right denotes Chern-number from lower band to upper band and the digit with bar denotes negative Chern-number. The topological phase transition denoted by the lines red, blue and black boxed-lines. The red, blue and black denotes the band gap closing between upper, middle and lower pairs of bands respectively. The thermal conductivity is plotted in Fig.\ref{THE} along the purple lines in (c) and (d). The purple lines in (c) corresponds to $D=0.5$. The line in (d) corresponds to $D=0.2$ .  }
	\label{Phases}
\end{figure}

The interplay between competing Heisenberg and DM interactions, together with
geometric frustration and external magnetic field results in
topologically ordered energy eigenstates. As the relative strengths of the 
different competing interactions are varied, the energy levels shift and the bands
cross / touch in pairs at different points in the Brillouin zone. The accompanying 
phase transitions are topological in nature as they are characterized by the change 
in Chern number of the pair of bands involved.  
By identifying the state of the system with the band topology\cite{triangular}, we find a wide variety 
of topological phases in different parameter regimes, shown in Fig.\ref{Phases}. The four numbers in the figure represent the four Chern numbers from lower to upper magnon bands. The bar above the number denotes the negative Chern number. The color of the phase boundaries identify the pair of bands involved in the transition. Most strikingly, tuning the strength of the different components of the DMI 
and applied magnetic field over a small range result in multitude of topologically distinct
set of single magnon bands, even though the ground state remains the unaltered (canted flux state). While
this is driven by the non-coplanarity of the ground state spin configuration, the exact
mechanism of the change in geometry of the magnon bands, or their robustness against interaction effects is not clear.\cite{triangular}. 


Topological phase transition occurs due to band reopening after closing at the the high symmetry points $\Gamma$, X, M and points along line $\overline{\Gamma\text{M}}$ and $\overline{\Gamma\text{M'}}$. Except for the $\Gamma$-point, all other k-points in the Brillouin zone can be mapped into another k-point using two-fold rotational symmetry. Thus, the Chern number of the bands changes by $\pm 1$, if band touching happens at the $\Gamma$-point. It is noticeable that this kind of phase transition happen in the upper right region of Fig.\ref{Phases}(b).  Otherwise, the the Chern numbers changes by $\pm 2$, because accidental band touching take places at two points in the Brillouin zone due to two-fold rotational symmetry of the system. Most of the phase transition is associated with change in Chern number $\pm 2$ in Fig.\ref{Phases}.

\subsubsection{\textbf{Thermal Hall conductance and its derivative}}

\begin{figure}[H]
	\centering
		\includegraphics[width=0.5\textwidth]{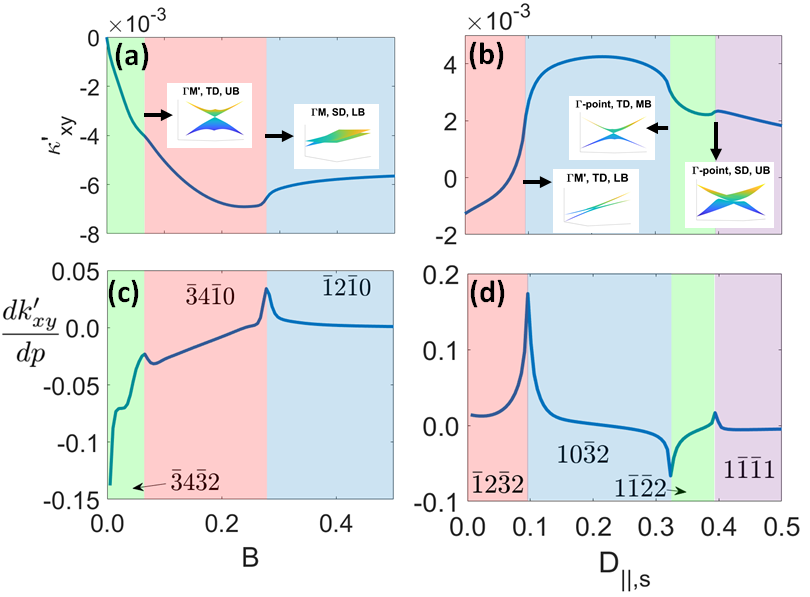}
	\caption{(color online) (a)-(b) The reduced thermal conductivity as a function of Magnetic field $B$ and DM-term $D_{||,s}$ along the purple lines in Fig.\ref{Phases}(d) and Fig.\ref{Phases}(b) respectively. Inset of the figure shows the band touching points at the boundaries of the topological region. The first, second and third letters denote the band touching point, type of band touching, and the energy wise ordering of bands.For example, "$\Gamma$M', TD, UB", band touches at $\Gamma$M' line , the type of band touching is tilted Dirac-type and Upper bands touch respectively. "SD" means the Semi-Dirac point and "UB" means Upper band touches and "MB" means middle band touches etc. The magnetic field is a reduced quantity and connected with experimental magnetic field as $h_z=\frac{BJ}{g\mu_B}$, where $g$ is Lande-g factor and $\mu_B$ is the Bohr magneton. (c)-(d) The derivative of the thermal Hall conductivity w.r.t. magnetic field and DM-term $D_{||,s}$  as a function of Magnetic field $B$ and DM-term $D_{||,s}$ respectively. The numbers in the  figures are Chern numbers with convention given in Fig.\ref{Phases}.}
	\label{THE}
\end{figure}

The non-trivial topology of magnon bands give rise to thermal Hall effect in the magnetic system.
The expression of reduced thermal Hall conductivity is given by\cite{THE1,THE2},
\begin{equation}
\kappa'_{xy}=\frac{\kappa_{xy}\hbar}{k_B}=\frac{T'}{(2\pi)^2}\sum_n \int_{BZ} c_2(\rho_{n,\bk}) \Omega^n_{xy}(\bk) d^2k,
\end{equation}
where, $\kappa_{xy}$ is the thermal Hall conductivity, $T'$ is the scaled temperature, $T'=k_B T$, and $\rho_{n,\bk}=1/(\exp(\epsilon^n(\bk)/T')-1)$ is the Bose-Einstein distribution function with $\epsilon^n(\bk)$ as the energy of the $n$-th magnon band at $\bk$-point in Brillouin zone. The reduced temperature $T'$, magnon energies $\epsilon^n(\bk)$ as well as reduced-thermal Hall conductivity are normalized in unit of $JS$. In Fig.\ref{THE}(a) and \ref{THE}(b), the results for the reduced thermal Hall conductivity is plotted as a function of Magnetic field $B_z$ and $D_{||,s}$ respectively, along the purple lines in Fig.\ref{Phases}(b) and Fig.\ref{Phases}(d). The different coloured regions in Fig\ref{THE}(a) and \ref{THE}(b) denote distinct topological regions along the purple lines in Fig.\ref{Phases}(d) and Fig.\ref{Phases}(b) respectively. At the boundary of the topological regions the band gap closes as shown in the inset of the figures. Generically, band closing occurs at the Dirac point, but sometime a semi-Dirac point is encountered. The type of semi-Dirac point at the boundary between green and purple topological regions in Fig.\ref{THE}(b) is also reported in the reference Ref.[\onlinecite{SemiDirac}].
Figs.\ref{THE}(c) and \ref{THE}(d) present the derivative of reduced thermal Hall conductivity  with respect to magnetic field $B$ and DMI $D_{||,s}$ as a function of $B$ and the  
symmetric component of the in-plane DMI $D_{||,s}$. At the boundary between two distinct topological phases, the derivative in the thermal Hall conductance has a logarithmic divergence. The origin of the divergence on the basis of Weyl-point also discussed in Ref.[\onlinecite{NonCoplanar2}]. It is observed from the figures that the logarithmic divergence is universal and independent of the type of band touching. We also have been shown analytically in Appendix\ref{AppendixF} that the nature of divergence is same for tilted Dirac point and Semi-Dirac point. Furthermore, it also can be seen from the figure that the peak height of divergence grows faster, if the band touching happens at the lower pair of bands,
due to the larger contribution to the thermal Hall conductivity from magnons in the lower bands. 
Finally, the sign of the divergence is positive (negative) if Chern number of lower band increases
(decreases) at the topological phase transition.

\begin{figure}[H]
	\centering
		\includegraphics[width=0.5\textwidth]{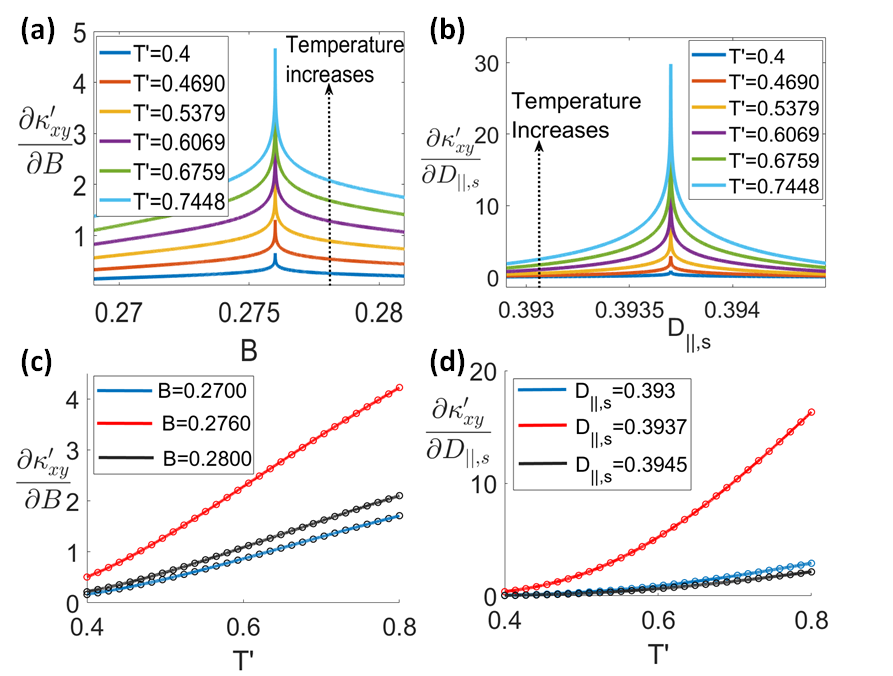}
	\caption{(color online) (a)-(b) The derivative of conductivity as a function of parameters $B$ and $D_{||,s}$ respectively, for different temperature. The method of plotting is described in the main text. (c)-(d) The derivative of conductance as a function of temperature is fitted using Eq.\ref{TemperatureDependenceEq}, at different $B$ and $D_{||,s}$ respectively. The circled points correspond to the calculated derivative of thermal Hall conductivity and the straight line represents the fitted curve. The fitting parameters $\left\lbrace A, \epsilon_0\right\rbrace$ for blue, red and black curves in (c) are $\left\lbrace 4.6886, 2.9243\right\rbrace$, $\left\lbrace 10.4323, 2.7539\right\rbrace$ and $\left\lbrace 5.5643, 2.8666\right\rbrace$. The fitting parameters $\left\lbrace A, \epsilon_0\right\rbrace$ for blue, red and black curves in (d) are $\left\lbrace 20.0830, 4.2135\right\rbrace$, $\left\lbrace 103.5474, 4.1025\right\rbrace$ and $\left\lbrace 15.4258, 4.2656\right\rbrace$.}
	\label{TemperatureDependenceFig}
\end{figure}

We found that the logarithmic divergence follows simple analytical expression as a function of temperature given as,
\begin{equation}
\frac{\partial \kappa'_{xy}}{\partial p}= A\left[\ln\left(\frac{1+\rho_0}{\rho_0}\right)\right]^2 \exp(\frac{\epsilon_0}{T})\rho_0^2,
\label{TemperatureDependenceEq}
\end{equation}
where, $\epsilon_0$ is the band touching point during topological phase transition, $\rho_0=1/(\exp(\epsilon_0/T)-1)$, $A$ is a constant independent of temperature and proportional to $\log(p)$, $p$ is a parameter of system(e.g. magnetic field etc.). Moreover, at a temperature, lower compared with the energy of band touching point, the equation \ref{TemperatureDependenceEq} transforms into,
\begin{equation}
\frac{\partial \kappa'_{xy}}{\partial p}=A\epsilon_0^2\exp(\frac{-\epsilon_0}{T}).
\end{equation}

To demonstrate the validity of Eq.\ref{TemperatureDependenceEq}, we have chosen the topological phase transition points near $B=0.276$ and $D_{||,s}=0.3937$ of Fig.\ref{THE}(c) and Fig.\ref{THE}(d) respectively. The numerically calculation Berry-curvature around the transition point is computationally expensive and in-accurate. So, to correctly calculate the derivative of thermal Hall-conductivity, first the thermal Hall conductivity has been calculated near the transition point and fitted using the expression,
\begin{equation}
\kappa'_{xy}=m\ln(|p-p_0|)+m_0+m_1(p-p_0)+m_2(p-p_0)^2+m_3(p-p_0)^3,
\end{equation}
where $p_0$ is the critical point. Then the derivative of the expression has been plotted and shown in Fig.\ref{TemperatureDependenceFig}(a) and Fig.\ref{TemperatureDependenceFig}(b). The divergent peak and nearby points increase with the temperature. In Fig.\ref{TemperatureDependenceFig}(c) and \ref{TemperatureDependenceFig}(d),  the derivative of conductivity is plotted and fitted as a function of temperature using the Eq.\ref{TemperatureDependenceEq}, considering $A$ and $\epsilon_0$ as fitting parameter. The band touches at $\epsilon_0=2.65$ for the phase transition at $B=0.276$ in Fig.\ref{TemperatureDependenceFig}(a) and at $\epsilon_0=4.079$ for the phase transition at $D_{||,s}=0.3937$ in Fig.\ref{TemperatureDependenceFig}(b). The values of fitting parameter $\epsilon_0$ described in Fig.\ref{TemperatureDependenceFig}(c) and Fig.\ref{TemperatureDependenceFig}(d) is quite near the band touching points.

\section{\label{sec5}Material Realizaion}
\begin{figure}[H]
	\centering
		\includegraphics[width=0.3\textwidth]{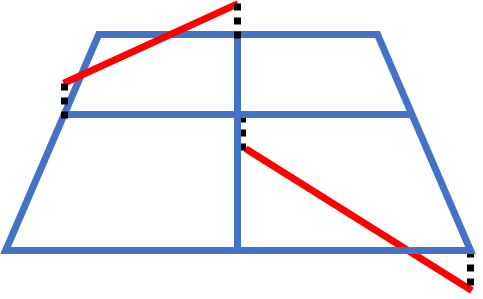}
	\caption{The symmetry of the Shastry-Sutherland lattice corresponding to the allowed DM-interactions in Fig.\ref{lattice}}
	\label{Buckled}
\end{figure}
We have carefully restricted our choice of Hamiltonian parameters to realistic 
ranges. Nearly equal Heisenberg exchange interactions on the diagonal and axial 
bonds($J\approx J'$) in SS-lattice is observed in the rare earth tetraborides
(\ce{RB4}, R=Er,Tm)\cite{RB41,RB42,RB43}. The nature of DM-interaction for the 
study(Fig.\ref{lattice}(b)) is based on the symmetry of \ce{CuO} layers of 
the canonical Shastry Sutherland compound, \ce{SrCu2(BO3)2} in the low temperature phase\cite{structure1, structure2, structure3}, where the bonds along the two
inequivalent diagonals of the unit cell are shifted out of the plane in opposite
directions as shown in Fig.\ref{Buckled}. Although the nature  and strength of DM-interaction in the native rare earth tetraborides has not been investigated experimentally, 
the possibility of forming Van der Waals heterostructures with heavy metals offers
the ability to induce and tune DM interactions over a extended range. The presence of
geometric frustration reduces the critical $D_\perp$ to achieve flux-state in SS-lattice,
facilitating the realization of the canted flux state and associated topological magnon
excitations in these materials. Finally, DM interactions can also be induced
by incident circularly polarized optical wave\cite{Floquet}, further enhancing the
possibility .


\section{\label{sec6}Conclusion}
In conclusion we consider the flux state of Shastry-Sutherland lattice and showed that in presence of in plane DMI and magnetic field, the system gives rise to non-trivial topological magnon bands. The canted flux state is a non-coplanar spin structure. This leads to a various topologically distinct magnon band structure. Again, we observed the nature of first derivative of thermal Hall conductance is logarithmic divergent at the topological phase transition, independent of the type of band touching. We have presented a simple temperature dependent parametric relation for the thermal Hall conductance, which might be useful to extract the energy of band touching during topological phase transition. Finally, we have suggested an experimental realization of the model studied. In the present work, we have assumed a dilute gas of magnons without any interaction. At finite temperatures, as the density of thermally excited magnons increase, effects of interaction gain importance. Interaction between magnons further re-normalizes the bands and impart a finite life-time, which in turn can change the topological phase diagram obtained in this study. The study of the topological magnon bands for this model in presence of interaction is planned for the future.

Financial support from the Ministry of Education, Singapore, in the form
of grant MOE2016-T2-1-065 is gratefully acknowledged.

\appendix

\section{Symmetry protected degeneracy in canted-flux state}
\label{AppendixC}

The symmetry protection of the band sticking at X-point in the Brillouin zone in the canted flux state without magnetic field has been explained using Herring’s method\cite{Herring,nonsymmorphic}.

The symmetry operators which keep the X-point in the Brillouin invariant or change it by a reciprocal lattice vector are,
\begin{align*}
&\left\lbrace M_2\middle| \boldsymbol{\delta}_2\right\rbrace,\;\left\lbrace M_1\middle|\boldsymbol{\delta}_1 \right\rbrace,
\\
&\left\lbrace C_{2z}\middle|\boldsymbol{\delta}_{y}\right\rbrace,
\end{align*}  
  where,
  \begin{align*}
 & M_2=\tilde{M}_2 e^{i\pi\hat{S}_z} e^{i\pi\hat{S}_1},\; M_1= \tilde{M}_1 e^{i\pi\hat{S}_z} e^{i\pi\hat{S}_2},\\
  & \;C_{2z}=\tilde{C}_{2z} e^{i\pi\hat{S}_z}e^{i\pi\hat{S}_1}e^{i\pi\hat{S}_z}e^{i\pi\hat{S}_2}.
\end{align*}   
The symmetry operators $\tilde{M}_2$, $\tilde{M}_1$ and $\tilde{C}_{2z}$ are mirror reflection  along axis-$P_1$, reflection along axis-$P_2$ and twofold rotation around z-axis at the sublattice-c respectively(Fig.\ref{FluxState}(a)).
 
The set of translational operators makes the invariant subgroup,
\begin{equation}
\pazocal{T}_\bold{k}=\left\lbrace E\middle| m\boldsymbol{\alpha}_1+n\boldsymbol{\alpha}_2\right\rbrace,\quad m\in \text{even},\; n\in \text{integer},
\end{equation}
where the translational operator follows the constraint $exp(i\bold{k}\cdot\bold{t})=1$ at X-point($\bold{t}=\left(\frac{\pi}{a},0\right)$).

Then the factor group $\pazocal{G}_\bold{k}/\pazocal{T}_\bold{k}$ can be obtained by deriving the coset of invariant subgroup $\pazocal{T}_\bold{k}$, where $\pazocal{G}_\bold{k}$ is the symmetry operators which keeps the X-point invariant or change it by reciprocal lattice vector. The factor group $\pazocal{G}_\bold{k}/\pazocal{T}_\bold{k}$ is given by, 

\begin{align*}
&e=\squeeze{E}{\mn}, \quad e'=\squeeze{E}{\pq} \\
&m_2=\squeeze{M_2}{\mn+\boldsymbol{\delta}_1}, \quad m'_2=\squeeze{M_2}{\pq+\boldsymbol{\delta}_1}\\
&m_1=\squeeze{M_1}{\mn+\boldsymbol{\delta}_2}, \quad m'_1=\squeeze{M_1}{\pq+\boldsymbol{\delta}_2}\\
&c_2=\squeeze{C_{2z}}{\mn+\boldsymbol{\delta}_x}, \quad c'_2=\squeeze{C'_{2z}}{\pq+\boldsymbol{\delta}_x},
\end{align*} 
where $m\in even$, $p\in odd$ and $n,q\in integer$.

Next, we have derived the character table using package named "GAP"\cite{GAP4}, and the derived character table is given by,
\begin{table}[H]
\centering
\caption{Character Table of unitary subgroup}
 \begin{tabular}{||c c c c c c||} 
 \hline
 & $e$ & $e'$ & $\squeezeA{m_2}{m'_2}$ & $\squeezeA{m_1}{m'_1}$ & $\squeezeA{c_2}{c'_2}$ \\ [0.5ex] 
 \hline\hline
$\Gamma_1$ & 1 &  1 &  1 &  1 &  1 \\ 
$\Gamma_2$ & 1 &  1 & -1 & -1 &  1 \\ 
$\Gamma_3$ & 1 &  1 &  1 & -1 & -1 \\ 
$\Gamma_4$ & 1 &  1 & -1 &  1 & -1 \\ 
$\Gamma_5$ & 2 & -2 & 0 & 0 & 0 \\ 
 \hline
\end{tabular}
\label{TableI}
\end{table}

The only valid representation is $\Gamma_5$, since the translational operators $e=\squeeze{E}{\pq}$ should follow the relation,
\begin{align*}
exp(i\bold{k}\cdot\bold{t})=-\pazocal{I},
\end{align*}
where, $\pazocal{I}$ is the identity matrix with a dimension equals to the dimension of representation.

Because  the only valid representation is $\Gamma_5$ which is two dimensional, the bands are doubly degenerate at the X-point.

In the similar manner, it can be shown that the degeneracy at the $\Gamma$ and M-points are also symmetry protected. 

\section{Magnon Hamiltonian and diagonalization}
\label{AppendixD}
The quadratic magnon Hamiltonian is given by,
\begin{align}
\pazocal{H}=S&\sum_{\bk}\left[Q_\bk(\done)\capdagk{a}\capk{c}+Q_\bk(-\dtwo)^*\capdagk{a}\capk{d}\right.\nonumber\\
&+Q_\bk(\dtwo)^*\capdagk{b}\capk{c}+Q_\bk(-\done)\capdagk{b}\capk{d}
\nonumber
\\
&-\frac{1}{2}(J+\epsilon_{A})\expmk{x}\capdagk{a}\capk{b}\nonumber \\
&\left.-\frac{1}{2}(J-\epsilon_{B})\expk{y}\capdagk{c}\capk{d})\right]+\mathtt{h.c.}
\nonumber
\\
+S&\sum_{\bk}\left[P_\bk(\done)^*\capk{a}\capmk{c}+P_\bk(-\dtwo)\capk{a}\capmk{d}\right.\nonumber\\
&+P_\bk(\dtwo)\capk{b}\capmk{c}+P_\bk(-\done)^*\capk{b}\capmk{d}
\nonumber
\\
&+\frac{1}{2}\left(J-\epsilon_{A}\right)\expk{x}\capk{a}\capmk{b}\nonumber \\
&\left.+\frac{1}{2}\left(J+\epsilon_{B}\right)\expmk{y}\capk{c}\capmk{d}\right]+\mathtt{H.c.}
\nonumber
\\
+S&\sum_{\bk}\left[\left(2E-\epsilon_{A}+\frac{B}{S}\cos(\theta_1)\right)\capdagk{a}\capk{a}\right.\nonumber\\
&+\left(2E-\epsilon_{A}+\frac{B}{S}\cos(\theta_1)\right)\capdagk{b}\capk{b}
\nonumber
\\
&+\left(2E+\epsilon_{B}+\frac{B}{S}\cos(\theta_2)\right)\capdagk{c}\capk{c}\nonumber\\
&\left.+\left(2E+\epsilon_{B}+\frac{B}{S}\cos(\theta_2)\right)\capdagk{d}\capk{d}\right],
\label{eq8}
\end{align}
where,
\begin{align}
E&=-2 J' C^X_\theta+2 D_\perp S^X_\theta -D_{||,ns}(\zeta_{\theta_1}-\zeta_{\theta_2})\nonumber\\&\quad-D_{||,s}(\zeta_{\theta_2}-\zeta_{\theta_1})
\nonumber\\
\epsilon_{A}&=J\cos(2 \theta_1)+D\sin(2 \theta_1)
\nonumber \\
\epsilon_{B}&=-J\cos(2\theta_2)+D\sin(2\theta_2)
\nonumber\\
Q_\bk(\delta)&=\lambda_{J'}^+(\bk,\delta)-\lambda_{D_\perp}^+(\bk,\delta)\nonumber\\
&-D_{||,ns}(\xi_{\theta_1}\expk{\delta}-\xi_{\theta_2}\expmk{\delta})\nonumber\\
&+D_{||,s}(\xi_{\theta_1}\expmk{\delta}-\xi_{\theta_2}\expk{\delta})
\nonumber\\
P_\bk(\delta)&=\lambda_{J'}^-(\bk,\delta)-\lambda_{D_\perp}^-(\bk,\delta)\nonumber\\
&-D_{||,ns}(\xi_{\theta_1}^*\expk{\delta}-\xi_{\theta_2}\expmk{\delta})\nonumber\\
&+D_{||,s}(\xi_{\theta_1}^*\expmk{\delta}-\xi_{\theta_2}\expk{\delta})
\nonumber\\
\nonumber\\
\lambda_{J'}^{\pm}(\bk,\delta)&=J'(i C^{\pm}_\theta+S^X_\theta)\cos(\bk\cdot\delta)
\nonumber\\
\lambda_{D_\perp}^{\pm}&=D_\perp (C^X_\theta \pm 1) \cos(\bk\cdot\delta)
\nonumber \\
\nonumber \\
\end{align}
\begin{align}
\xi_{\theta_i}&=\frac{i}{2}\sin(\theta_i)+\frac{1}{2}\cos(\theta_i)\sin(\theta_{\bar i})
\nonumber \\
\zeta_{\theta_i}&=\sin(\theta_i)\cos(\theta_{\bar i}),\;\; (i,\bar{i}) = (1,2)\; \text{or}\; (2,1)
\nonumber \\
C^X_\theta &=\cos(\theta_1)\cos(\theta_2)
\nonumber\\
C^{\pm}_\theta &=\cos(\theta_1)\pm\cos(\theta_2)
\nonumber \\
S^X_\theta &=\sin(\theta_1)\sin(\theta_2)
\nonumber\\
\label{eq9}
\end{align}

The Hamiltonian can be re-written in the form of Bogoliubov de Gennes matrix,
\begin{equation}
\pazocal{H}=\sum_\bk \Psi^\dagger(\bk)\pazocal{H}(\bk) \Psi(\bk),
\end{equation}
where $\Psi^\dagger(\bk)=(\hat{a}^\dagger_\bk,\hat{b}^\dagger_\bk,\hat{c}^\dagger_\bk,\hat{d}^\dagger_\bk,\hat{a}_{-\bk},\hat{b}_{-\bk},\hat{c}_{-\bk},\hat{d}_{-\bk})^T$. The matrix $\pazocal{H}(\bk)$ is diagonalized using the para-unitary transformation,
\begin{equation}
T^\dagger(\bold{k}) \pazocal{H}(\bold{k}) T(\bold{k})=\epsilon(\bold{k}),
\label{eqI}
\end{equation}
where,
\begin{equation}
\epsilon(\bold{k})=\begin{pmatrix}
E_1(\bold{k}) & & & &\\
& ... & & & &\\
& & E_{N}(\bold{k}) & & &\\
& & &  E_1(-\bold{k}) & &\\
& & & & ...&\\
& & & & & E_{N}(-\bold{k})
\end{pmatrix}.
\end{equation}
The para-unitary transformation preserves the Bosonic commutation relation of the eigenvectors obtained after diagonalization.

The para-unitary matrices follow the relation,
\begin{equation}
T^\dagger(\bold{k}) \sigma_3 T(\bold{k})=T(\bold{k}) \sigma_3 T^\dagger(\bold{k})=\sigma_3,
\label{eqIII}
\end{equation}
where $\sigma_3$ is the 3rd component of Pauli matrix. Using Eq.\ref{eqIII}, we can re-write Eq.\ref{eqI} as,
\begin{equation}
\sigma_3 \pazocal{H}(\bold{k}) T(\bold{k})=T(\bold{k})\sigma_3\epsilon(\bold{k})
\label{D7}
\end{equation}
So the problem of para-unitary transformation become the problem of diagonalization of the matrix $\sigma_3\pazocal{H}(\bold{k})$.

\section{Calculation of Chern number and Berry curvature}
\label{AppendixE}

The presence of non-zero Chern numbers of the bands denote non-trivial topology of the magnon bands. The expression of the Chern number for n-th band is given by,

\begin{equation}
C_n=\frac{1}{2\pi}\int_{\text{BZ}} \Omega^n_{xy}(\bk) dk_x dk_y,
\end{equation}
where the Berry curvature\cite{Berry},
\begin{equation}
\Omega^n_{xy}(\bk)=i\epsilon_{\mu\nu}\left[\sigma_3 \dXdY{T^\dagger_\bk}{k_\mu}\sigma_3\dXdY{T_\bk}{k_\nu}\right]_{nn} \qquad \left(n=1,2,3,...,2N\right),
\label{berry1}
\end{equation}
Where $N$ is the number of sublatices and $T_\bk$ is the  para-unitary matrix which diagonalizes the Bogoliubov Hamiltonian(See Appendix.\ref{AppendixD}). Each column of the para-unitary matrix $T_\bk$ corresponds to the wave-functions at $\bk$. The Eq.\ref{berry1} is not suitable for numerical calculation, since numerically Gauge degrees of freedom of the wave-function is not well controlled and so the derivative of the wave-function is ill defined. The Berry curvature in Eq.\ref{berry1} can be re-structured as(See Appendix.\ref{AppendixE}),
\begin{align}
& \Omega^n_{xy}(\bk)=\sum_{m\neq n} i (\sigma_3)_{nn}(\sigma_3)_{mm} \nonumber\\ &\left[\frac{\squeezeB{\uu{n}}{\dXdY{\pazocal{H}_\bk}{k_x}}{\uu{m}}\squeezeB{\uu{m}}{\dXdY{\pazocal{H}_\bk}{k_y}}{\uu{n}}-(k_x\leftrightarrow k_y)}{((\sigma_3\epsilon_\bk)_{nn}-(\sigma_3\epsilon_\bk)_{mm})^2}\right],
\end{align}
where, $\left|u^m(\bk)\right\rangle$ is the m-th column of the para-unitary matrix $T(\bk)$ and $\left\langle u^m(\bk)\right|$ is the m-th row of the para-unitary matrix $T^\dagger(\bk)$.

To calculate the non-Abelian Berry curvature, we followed the method described in Ref.\onlinecite{Hatsugai}. For the sake of completeness, here we describe the calculation briefly. The expression of Chern number is given by,
\begin{equation}
\tilde{F}_{12}(k_l)=\frac{1}{i}\ln(U_1(k_l)U_2(k_l+\hat{1})U_1(k_l+\hat{2})^{-1}U_2(k_l)^{-1}),
\end{equation}
where, $k_l$ is the discrete k-point in the Brillouin zone and if $N_i$ is the number of lattice points in the $i$-th direction(where $i=x$ or $y$) then, $l=1,2,...,N_1N_2$.  Again $k_l+\hat{i}$ denotes the k-point next to $k_l$ along the $i$-th direction. Again,
\begin{equation}
    U_\mu(k_l)=\text{det}(\psi^\dagger(k_l)\psi(k_l+\hat{\mu}))/N_\mu(k_l),
\end{equation}
where, $N_\mu(k_l)=|\text{det}(\psi^\dagger(k_l)\psi(k_l+\hat{\mu}))|$ and $\psi(k_l)=(T_\bk)_{mn}$. The indices $m$ and $n$ denote the band-indices for which the non-Abelian Berry-curvature is calculated.

\section{Logarithmic divergence in derivative of thermal Hall conductivity near phase transition}
\label{AppendixF}
\subsection{Tilted Dirac point or Generalized Weyl point}
 \label{TiltedDirac}
 The Hamiltonian corresponds to the tilted Dirac point or generalized Weyl point is given by,
\begin{equation}
\hat{H}=w^0_x k_x \mathbb{I}+w^0_y k_y \mathbb{I}+w_x k_x \sigma_x + w_y k_y \sigma_y + p \sigma_z +\epsilon\mathbb{I},
\label{Hamiltonian1}
\end{equation} 
where $k_x$, $k_y$ are the momentum with respect to the band touching point. $\sigma_i$ ($i=x,y,z$) are the Pauli's matrices. $p$ is a perturbation to open the gap. $\epsilon$ denotes the energy of band touching point. \textbf{The Hamiltonian is general Hamiltonian for any linear dispersions}.

The energies correspond to the Hamiltonian,
\begin{align}
E^{\pm}(\bk)&=\epsilon+k_x w^0_x +k_y w^0_y \pm \sqrt{p^2+w^2_x k^2_x+w^2_y k^2_y}\nonumber\\
&=\epsilon_{0\bk} \pm \omega_{\bk},
\label{Energy1}
\end{align}  
where $\epsilon_{0\bk}=\epsilon_0+k_x w^0_x +k_y w^0_y $ and $\omega_{\bk}=\sqrt{p^2+w^2_x k^2_x+w^2_y k^2_y}$
 the corresponding eigenvectors are,
 \begin{equation}
 x^\pm=\frac{1}{N_{\pm}}\begin{pmatrix}
 \frac{p\pm \sqrt{p^2+w^2_x k^2_x+w^2_y k^2_y}}{w_x k_x+ i w_y k_y} \\
 1
 \end{pmatrix},
 \label{eigenvector1}
 \end{equation}
 
 where $N_{\pm}$ are the normalization constant. 
 
 The expression of the Berry curvature of the lower band is given by,
 \begin{align}
 \Omega_1 &=\frac{1}{(E^+-E^-)^2}\text{Im}\left(\squeezeD{x_1}{\dXdY{\pazocal{H}}{k_x}}{x_2}\squeezeD{x_2}{\dXdY{\pazocal{H}}{k_y}}{x_1}\right),
 \label{BerryCurvature1}
 \end{align}
 where Im$(z)$ denotes imaginary part of $z$.
 
 Using Eq.\ref{Hamiltonian1}, Eq.\ref{Energy1} and Eq.\ref{eigenvector1},
 \begin{equation}
 \squeezeD{x_1}{\dXdY{\pazocal{H}}{k_x}}{x_2}=\frac{2w_x}{\sqrt{N_1 N_2}} \left[ \frac{w_x k_x p-i w_y k_y \sqrt{p^2 w^2_x k^2_x+w^2_y k^2_y}}{w^2_x k^2_x+w^2_y k^2_y}\right]
 \label{auxilary1a}
 \end{equation}

Similarly,
 \begin{equation}
	 \squeezeD{x_2}{\dXdY{\pazocal{H}}{k_y}}{x_1}=\frac{2w_y}{\sqrt{N_1 N_2}} \left[ \frac{w_y k_y p-i w_x k_x \sqrt{p^2 w^2_x k^2_x+w^2_y k^2_y}}{w^2_x k^2_x+w^2_y k^2_y}\right] 
	 \label{auxilary1b}
 \end{equation}
 
 Using Eq.\ref{auxilary1a} and Eq.\ref{auxilary1b} in Eq. \ref{BerryCurvature1}, the Berry curvature of the lower band,
 \begin{equation}
 \Omega_1=-\frac{2w_x w_y p}{(p^2+w^2_x k^2_x + w^2_y k^2_y)^{3/2}}
 \end{equation}
 
 Similarly, for the upper band,
  \begin{equation}
 \Omega_2=+\frac{2w_x w_y p}{(p^2+w^2_x k^2_x + w^2_y k^2_y)^{3/2}}
 \end{equation}
 
The Thermal Hall conductivity expression is given by,
\begin{equation}
\kappa_{xy}=\frac{k_B^2 T}{(2\pi)^2 \hbar} \sum_n \int_{BZ} c_2(\rho_{n\bk}) \Omega_n(\bk) d^2k, 
\label{ThermalHallEffect1}
\end{equation}

\begin{align}
\sum_n \Omega_n(\bk) c_2(\rho_{n\bk})&=\frac{2w_x w_y p}{(p^2+w^2_x k^2_x + w^2_y k^2_y)^{3/2}} \left[ c_2(\rho_{2\bk})-c_2(\rho_{1\bk})\right] \nonumber\\
&\propto \frac{2w_x w_y p}{(p^2+w^2_x k^2_x + w^2_y k^2_y)^{3/2}} \omega_\bk\quad \nonumber \\  \left[ where,c_2(\epsilon_{0\bk}+\right. & \left.\omega_{\bk})-c_2(\epsilon_{0\bk}-\omega_{\bk})\propto \omega_\bk \text{for small $p$ and $\bk$ } \right] \nonumber\\
&\propto \frac{p}{(p^2+w^2_x k^2_x + w^2_y k^2_y)}
\end{align}
 
 From Eq.\ref{ThermalHallEffect1}, integrating around the band touching points $k=\sqrt{w^2_x k^2_x + w^2_y k^2_y}<k_c$, we get,
 \begin{align}
 \dXdY{\kappa_{xy}}{p} &\propto \ddY{p}\int_{k<k_c}  \frac{p}{(p^2+w^2_x k^2_x + w^2_y k^2_y)} d^2k  \nonumber \\
 &\propto \int_{k<k_c} \frac{2\pi k dk}{(p^2+k^2)} - \int_{k<k_c} \frac{4p^2\pi k dk}{(p^2+k^2)^2}\nonumber \\
 &\propto \pi \ln\left(\frac{p^2+k^2_c}{p^2}\right) + 2p^2\pi  \left[\frac{1}{p^2+k^2_c}-\frac{1}{p^2}\right]\nonumber\\
 &\propto \ln(|p|) \quad \left[\text{Near phase transtion $k_c\gg p$}\right]
 \end{align}

 \subsection{Semi-Dirac point}
The dispersion of semi-Dirac point,
\begin{equation}
E^{\pm}(\bk)=\pm \sqrt{(w^2_x k^2_x)^2+w^2_yq^2_y}
\label{Energy2}
\end{equation}

  Out of many possibilities, two different possible Hamiltonians are,
 \begin{equation*}
 \pazocal{H}_1=\begin{pmatrix}
 0 & w_x q_x  +i w_y q_y \\
 w_x q_x-i  w_y q_y  & 0
 \end{pmatrix}
\end{equation*}

 \begin{equation}   
 \pazocal{H}_2=\begin{pmatrix}
w_x q_x & i w_y q_y\\
-i w_y q_y & -w_x q_x 
 \end{pmatrix},
 \label{Hamiltonian2}
 \end{equation}
 
The Berry curvatures of the both Hamiltonian is given by,
\begin{equation}
\Omega^\pm_1=\pm \frac{4pq_x w_x w_y}{(p^2+w^4_x q^4_x +w^2_y q^2_y)^{3/2}}, \Omega^\pm_2=0,
\end{equation}  
 Thus, the second Hamiltonian in Eq.\ref{Hamiltonian2} does not produce any Berry curvature and so topological phase transition can not happen with this kind of Hamiltonian. Similarly, it can be proved that,
\begin{equation}
 \dXdY{\kappa_{xy}}{p} \propto \ln(|p|)
 \end{equation}  

Particularly, in our study, we got semi-Dirac points with dispersion,
 \begin{equation}
 E^\pm(\bk)=w_0^2 k_y^2\pm \sqrt{w_x^2 k_x^2 + w_y^2 k_y^4}
 \end{equation}
Again particularly for phase transition point, near $D_{||,s}=0.4$ in Fig.\ref{THE}(d) is associated with Semi-Dirac dispersion with $w_0=w_y$, which is also encountered in reference\cite{SemiDirac}. 

\section{Temperature dependence of derivative of thermal Hall conductivity near phase transition}
\label{AppendixG}
The reduced thermal Hall conductivity is given as,
\begin{equation}
\kappa'_{xy}=\frac{\kappa_{xy}\hbar}{k_B}=\frac{T'}{(2\pi)^2}\sum_n \int_{BZ} c_2(\rho_{n,\bk}) \Omega^n_{xy}(\bk) d^2k.
\end{equation}

Near a topological phase transition, the derivative of the of Thermal hall conductivity contributes to a logarithmic divergence. So, the main contribution of derivative of thermal Hall conductance comes from the band touching point at energy $\epsilon_0$. Near this point at phase transition the Berry curvature is equal and opposite for the two bands$(\Omega^1_{xy}=-\Omega^2_{xy})$. Near the band touching point,

\begin{align}
\frac{\partial \kappa'_{xy}}{\partial p} &=\frac{T'}{2\pi^2}\frac{\partial}{\partial p} \left[ \Omega^1_{xy}(\bk) \left( c_2(\rho(\epsilon_0+\epsilon_{\bk})-c_2(\rho(\epsilon_0-\epsilon_{\bk}))\right)\right]\nonumber\\
&=\frac{T'}{2\pi^2}\frac{\partial}{\partial p}\left[ \Omega^1_{xy}(\bk)\;\left( 2\left.\frac{dc_2}{d\rho}\right|_{\rho_0} \left.\frac{d\rho}{d\epsilon}\right|_{\epsilon_0}\right)\epsilon_{\bk}\right]\nonumber\\
&=A\rho_0^2 \exp(\frac{\epsilon_0}{T'}) \left[\ln\left(\frac{1+\rho_0}{\rho_0}\right)\right]^2,\;
\end{align}
where, $\rho_0=1/(\exp(\frac{\epsilon_0}{T'})-1)$ and $A$ is a temperature independent parameter.

\bibliographystyle{apsrev4-1}

\end{document}